# Using Deep Learning and Explainable Artificial Intelligence in Patients' Choices of Hospital Levels

Lichin Chen[1], Yu Tsao[1], Ji-Tian Sheu[2]

[1] Research Center for Information Technology Innovation, Academia Sinica, Taipei, Taiwan

[2] Department of Health Care Management, Chang Gung University, Taoyuan, Taiwan

Abstract

**Background and Objective:** In countries that enabled patients to choose their own providers, a common problem is that the patients did not make rational decisions, and hence, fail to use healthcare resources efficiently. This might cause problems such as overwhelming tertiary facilities with mild condition patients, thus limiting their capacity of treating acute and critical patients. **Methods:** To address such maldistributed patient volume, it is essential to oversee patients' choices before further evaluation of a policy or resource allocation. This study used nationwide insurance data, accumulated possible features discussed in existing literature, and used a deep neural network to predict the patients' choices of hospital levels. This study also used explainable artificial intelligence methods to interpret the contribution of features for the general public and individuals. In addition, we explored the effectiveness of changing data representations by comparing the performance of the model trained with and without the preprocessing by an autoencoder. **Results:** The results showed that the choice of patients was highly imbalanced. However, the model was still able to predict with high area under the receiver operating characteristics curve (AUC) (0.90), accuracy (0.90), sensitivity (0.94), and specificity (0.97) without initially finding the best-fit features. Generally, social approval of the provider by the general public (positive or negative) and the number of practicing physicians serving per ten thousand people of the located area are listed as the top effecting features. The changing data representation had a positive effect on the prediction improvement. All the performance indicators increased after applying the autoencoder. **Conclusions:** Deep learning methods can process highly imbalanced data and achieve high accuracy. The effecting features affect the general public and individuals differently. Addressing the sparsity and discrete nature of insurance data leads to better prediction. Applications using deep learning technology are promising in health policy making. More work is required to interpret models and practice implementation.

# 1. Introduction

While ensuring the accessibility to care, some countries applied a "gate-keeping" strategy and others empowered patients with the freedom to choose their own providers. In the gate-keeping strategy, the patients first visit general practitioners (GP) for medical advice, and the GP decides whether to refer the patient to a secondary or tertiary institution. The intention of the gate-keeping strategy is to enhance efficiency, regulate cost, and reduce wait time for secondary care. However, there are some limitations to this strategy such as GPs limiting the possibility of specialists responding to patient and market demands [1, 2]; further, the subsequent delayed referral can cause other problems [3]. Some countries have reformed their strategy to offer patients with the freedom to choose providers. Patients are to "vote with their feet" and choose health providers who fit their preferences and needs [4]. This strategy empowers the patients by prompting providers to compete for patients through a customer–market mechanism, such as improving their services, such as care quality, efficiency, and wait time [2, 5]. However, such expectations has preconditions. Patients are expected to make their choices based on sufficient information and rational decision [2, 5], which is commonly not the case. Other studies report that patients have shown inadequate ability to use comparative information during provider selection [2, 6]. Past studies indicate that the patients' choice is a complex interrelationship between the characteristics of the patients, the providers, and the incident itself [2]. The decision may differ based on the characteristics of individuals, the characteristics of the provider, and the condition of the incidence, which makes it difficult to evaluate in advance.

However, researchers have developed several techniques to simulate and forecast the patient flow, patient volume, resource allocation, and patient choices. The gravity model, for instance, calculates the spatial interaction between a community and a hospital using population mass of the community, capacity and service mix of hospitals, and distances (or traveling time) [7, 8]. The aggregate hospital choice model (AHCM) is intended to model hospital choices through the market share. Based on historical data, AHCM uses time-series techniques to forecast future patient volumes [9]. Forecast and simulation techniques such as mean absolute percentage errors, autoregressive integrated moving average (ARIMA), seasonal ARIMA [10], and discrete event simulation models [11, 12] have also been used previously. However, these theories have strict preconditions and partially explain the choice scenario. Recently, deep learning methods have gained popularity. They capture the underlying pattern of data by transforming data into a more abstract matter and classifying them based on the latent distribution [13, 14]. The deep learning approach has been proven effective and has shown excellent performance in a wide variety of applications [13, 15, 16], such as disease risk forecasting [17], vital signs classification into physiological symptoms [18, 19], image classification for diagnosis [20, 21], text-based medical condition recognition [22, 23], and clinical event forecasting [24]. However, irrespective of the outstanding performance, the results of the deep learning technology remained a blackbox, which merely provided the results without reasons or any information that indicated how the conclusion was reached. This incapability of the deep learning method to gain trust and convince people to use its results limited its implementation in healthcare field. Meanwhile, some previous studies indicate that the existing hierarchical coding scheme of electronic health recodes is not sufficiently representative [25, 26]. It does not quantify the inherited similarity between concepts, and using deep learning to project discrete encodings into vector spaces may lead to a better analysis and prediction. While insurance data are the most commonly used data in policymaking, it is necessary to investigate the pattern of insurance data before applying deep learning.

This study accumulates the possible features that were previously mentioned in related work on the patients' choices and aims to use deep neural networks (DNNs) to generalize and predict the patients' choices. Focusing on the hospital levels of the patients' choices, this study used explainable artificial intelligence (XAI) methods to interpret the effecting features for the general public and individuals. In addition, this study explored the representations of insurance data by comparing the performance of model training with and without the preprocessing of changing data representations. The data used were the insurance data of the National Health Insurance (NHI) of Taiwan.

*1.1 Effecting features of patients' choices*

We characterized all the effecting features based on three main entities: the patient, the provider, and the incident. The characteristics of patients such as the age, gender, income, and previous medical experiences (positive and negative) were all possible features affecting the behavior of patients while accessing care [2, 6, 27]. Further, young people need relatively less care and females are more endurable than males; these facts along with the income of individuals affected the patients' willingness to visit a facility. Meanwhile, studies [28] indicated that the patients' satisfaction and loyalty were significantly related to the facility they would go to. When patients were satisfied with and loyal to a specific provider, they were less likely to change providers. Continuity of care was also considered an effecting feature, which could be indicated as the duration and frequency of visiting a provider (known as density); those who visited a provider regularly were less likely to change providers. The frequency of changing providers (known as dispersion) could also be an indicator of continuity of care.

The characteristics of a provider included the hospital reputation, hospital level, facilities at the hospital, and travel distances. Commonly, patients considered that institutes with higher levels (secondary and tertiary institutes) possess better equipment, more skillful physicians, and higher reputations. Some studies indicate that service quality affects the patients' demand [29], while others report that patients commonly neglect the quality indicators and prefer recommendations from associates [2]. Professionals such as GPs are tempted to decide based on feedback from patients and colleagues as well as their cooperation experiences with the department or hospital, rather than official information such as quality of service or wait time [6]. Some patients even decide based on review websites [4, 30], which is another form of social approval and recommendations from others. Studies [28] also indicate that the patients' choices may change according to the condition of the incident, for example, the severity of the condition, the complication of the disease, and the number of patients with chronic disease. The health status of the patients would affect their willingness to travel distance. Non-critical events and healthier patients would consider farther distance of travel and seek a second opinion. People would be tempted to go to emergency services as outpatient services are closed on weekends or holidays.

*1.2 Hospital choice of Taiwanese people*

People in Taiwan have the freedom to choose their own providers without the referral of a GP [31, 32]. Under universal coverage, people do not possess the knowledge of choosing providers rationally. Other than choosing physicians, people commonly consider the level of the institute. Primary care service in Taiwan are usually a physician of a certain specialty who owns a clinic and acts as a private practitioner [31, 33]. Hence, the primary care service commonly referred to clinics irrespective of specialty. A secondary and tertiary care often referred to regional/district hospitals and medical centers. People commonly consider that "large hospitals possess more skillful physicians and better medical facilities," resulting in patients, irrespective of their medical severity, swarming to medical centers [31, 34]. According to a public opinion poll conducted in 2019 [35], although 85.3% of the respondents agreed that for a mild condition the patient should go to the primary care service nearby instead of tertiary hospitals, 70% considered institutes with higher levels to possess better professional skills, and 49% expressed having confidence in determining the severity of their own condition. This phenomenon causes the recession of the primary care service and overcrowding of the tertiary care [33]. Consequently, tertiary facilities are overwhelmed with mild-conditioned patients and have limited capacity to treat acute and critical patients [31, 32]. This also induces an increase in the cost of treating mild conditions. The "hierarchization of services," highlighted for several years, refers to visiting appropriate medical resources according to needs rather than swarming to tertiary institutes. Approaches to ease such extreme unbalanced patient volume have been proposed, such as increasing copayments, strengthening referral mechanisms, limiting outpatient service volumes for tertiary facilities, and providing incentives for cooperation between different levels of providers. However, such a phenomenon still exists [32, 33]. To address the maldistribution of patient volume, it is essential to oversee the choice of patients before strategy planning. This study aims to support the evaluation by providing a sophisticated tool to predict the patients' choice and provide the effecting features for them to do so.

## 2. Material and Methods

The aim of this study is to predict the patients' choices of hospital levels and interpret the reasons for the prediction. In addition, this study demonstrates the effectiveness of changing the representation of insurance data. The data used were the insurance-claims data from the two million clinical declaration files and the Registry for Beneficiaries files from the Taiwan NHI research database (NHIRD), dated from January 1, 2008, to December 31, 2011. The data were originally sampled to ensure their representation of the population across Taiwan. The files included the demographic information and visiting records of outpatients and emergency settings. Some publicly announced data were added to enrich the records, including the "physician density" information that referred to the number of practicing physicians serving per ten thousand people in each region of Taiwan [36]. The national calendar was used to retrieve information on weekends and national holidays. Incomplete or questionable data, such as individuals without birth date or gender (or with two genders), records without a date, birth date later than the visit date, patients without any visiting records, patients without a primary diagnosis, and incomplete information of visited hospitals were excluded. Eighteen features were included in the analysis, which were characteristics of the patients, providers, and incidents. The following section will further outline the definition and calculation equation of the features. The targeted prediction outcome of this research is the four hospital levels, namely the medical center, regional hospital, district hospital, and clinic.

*2.1 Characteristics of the Patient*

The age, gender, low income (Yes/No), total number of visits, total number of diseases, total number of chronic diseases, and four continuity indicators were included as characteristics of the patient. The age was determined according to the date of the visit. The denotation of low income was the status identified when entering the insurance. The total number of visits was the number of visiting records during the study period. The total number of diseases and chronic diseases were identified using the encoded International Classification of Diseases (ICD) codes for each patient. The 4 continuity indicators captured the duration, density, dispersion, and continuity of an individual while accessing care [37-39]. To track duration and density of patients accessing care, the usual provider of care (UPC) and least usual provider of care (LUPC) were used to indicate the visiting ratio of medical institutes. The UPC represented the most frequently visited institute, and the LUPC represented the least frequently visited institute. The patients were required to visit the provider at least once to denote an institute as the LUPC. The sequential continuity of care index (SECOC) was used to calculate the change of providers, representing the dispersion of care. The Continuity of Care index (COCI) was a single indicator that represented the continuity of care for an individual. The calculation of the 4 continuity indicators are shown in equations (1), (2), (3), and (4), where $N$ represents the total number of visits, $n_i$ denotes the number of visits to the $i^{th}$ provider, $k$ is the number of providers once visited, and $C_j$ is denoted as 1 when the $j^{th}$ provider is the same as the $(j+1)^{th}$ provider (0 if not).

$$\text{UPC} = \max\left(\frac{n_i}{N}\right) \tag{1}$$

$$\text{LUPC} = \min\left(\frac{n_i}{N}\right) \tag{2}$$

$$\text{SECOC} = \frac{\sum_{j=1}^{N-1} C_j}{N-1} \tag{3}$$

$$\text{COCI} = \frac{(\sum_{i=1}^{k} n_i^2) - N}{N(N-1)} \tag{4}$$

*2.2 Characteristics of the Provider*

We characterized the providers with 3 indicators: the physician density of each region, the most frequent provider continuity (MFPC), and the least frequent provider continuity (LFPC). With reference to the patients' "vote with their feet" in choosing providers, we calculated the MFPC and LFPC to represent the patients' experiences and recommendations for each institute [40, 41]. The MFPC represented the frequency of being voted as the UPC, and the LFPC represented the frequency of being voted as the LUPC. Each patient could vote for only one MFPC and LFPC. The calculations are shown in equations (5) and (6), where *p* indicates the total number of patients, and $\text{UPC}_i$ and $\text{LUPC}_i$ indicate the $i^{th}$ patient who voted the provider as the UPC or LUPC, respectively.

$$\text{MFPC} = \sum_{i=1}^{p} \text{UPC}_i \tag{5}$$

$$\text{LFPC} = \sum_{i=1}^{p} \text{LUPC}_i \tag{6}$$

*2.3 Characteristics of the Incident*

The incident was characterized based on five encoded features. Through the encoded ICD codes and treatment codes of visiting records, we identified whether a surgery was involved (Yes/No), whether it was an emergency service (Yes/No), whether it was considered as a severe condition (Yes/No), whether the visit day was a work day (Yes/No), and the disease importance rate (DIR) of the target disease during that visit. The identification of surgeries and emergency services were based on the treatment codes defined by the NHIRD. The severity was defined based on emergency triage results. Triage results rank for level 1 to 3 (1 = resuscitation, 2 = emergency, and 3 = urgent) were listed as severe, and conditions that were included in the catastrophic illness announced by the NHI were also listed as severe. The date of the incident was distinguished as a work day or non-work day.

To capture whether the visit of the patient was a regular or singular event, we used the DIR to represent the importance of the target disease in that visit. The encoded primary diagnosis was identified as the target disease in that visit. The DIR represents the ratio of importance and is calculated as shown in equation (7), where $N$ indicates the total number of visits of the patient, and $d_i$ represents the total number of visits for disease $d_i$. We used only the primary diagnosis of the visit to identify the DIR.

$$\text{DIR} = \frac{d_i}{N} \tag{7}$$

*2.4 Deep Learning Framework*

This study used the DNN framework to train the patients' choices of hospital levels. DNN is a complex version of an artificial neural network (ANN) that contains multiple hidden layers [42, 43], where every neuron in layer $i$ is fully connected to every other neuron in layer $i + 1$. In a multi-layer neural network, each layer of the network is trained to produce a higher level of representation of the observed pattern. Every layer produces a representation of the input pattern that is more abstract than the previous layer by composing more nonlinear operations [13, 44]. The computation is shown in equation (8). Each hidden layer computes a weighted $w_{ij}$ and bias $b_{ij}$ of the output from the previous layer, followed by a nonlinear active function $\sigma$ that calculates the sum as outputs. The number of units in the previous layer is represented by $d$ and the output of the previous layer by $x_j$. Figure 1 demonstrates the DNN architecture.

$$\hat{y} = \sigma(\sum_{j=1}^{d} x_j w_{ij} + b_{ij}) \tag{8}$$

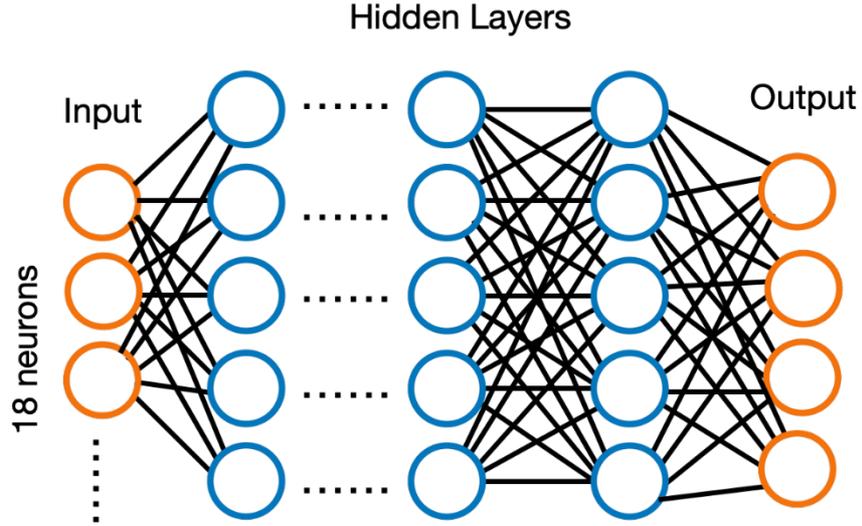

Figure 1. Deep Neural Network Architecture

To demonstrate the effect of changing data representations of insurance data, we designed a comparison. We used an autoencoder (AE) as the processor for the data representation change. AE is popular for processing scarce and noisy data [25, 26, 45]. It encoded the input into a lower dimension space $z$ and then decoded the representation by reconstructing an approximate input $\tilde{x}$. The goal of the reconstruction was to minimize the mean square error of $x$ and $\tilde{x}$. Equations (9) and (10) demonstrate the computation of the encoder and decoder, where $W$ and $W'$ denote the respective weights and $b$ and $b'$ denote the respective bias of the encoder and decoder. Figure 2 demonstrates the AE architecture.

$$z = s(W_x + b) \qquad (9)$$

$$\tilde{x} = s(W'_z + b') \qquad (10)$$

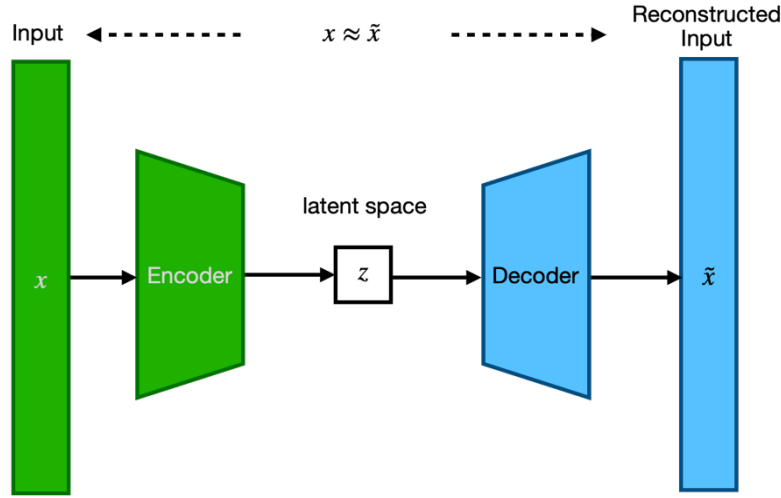

Figure 2. Autoencoder Architecture

XAI methods enhance the interpretability of machine learning models [46, 47]. This study adopted the shapley additive explanations (SHAP) [48]. SHAP combines the desirable characteristics of other interpretation frameworks, including local interpretable model-agnostic explanations (LIME) and deep learning important features (DeepLIFT). The SHAP value was computed using all combinations of the input, and the average marginal contribution of a feature value over all possible coalitions was calculated. SHAP has the ability to interpret models globally and locally, that is, to show the general effects of features on the whole population and individuals.

*2.5 Data processing flow*

All features were aggregated into a visit vector. Due to the imbalanced distribution of hospital levels, this study used a random undersampling strategy to sample the majority label and balance the training set. The model was trained on balanced data and tested on actual distributed data [49]. Meanwhile, to deal with numerical features with different scale levels, all the numerical values (including the age, number of diseases and chronic diseases, number of visits, number of votes as MFPC and LFPC, and physician density) were normalized between 0 and 1, and the categorical features were transformed into a one-hot/dummy encoding before analysis. Those indicators that were already ratio figures (values between 0 and 1) were used accordingly (including COCI, UPC, LUPC, SECOC, and DIR). The data were randomly split into training data (80%) and testing data (20%). A five-fold cross-validation training strategy was used. The data processing flow is shown in Figure 3.

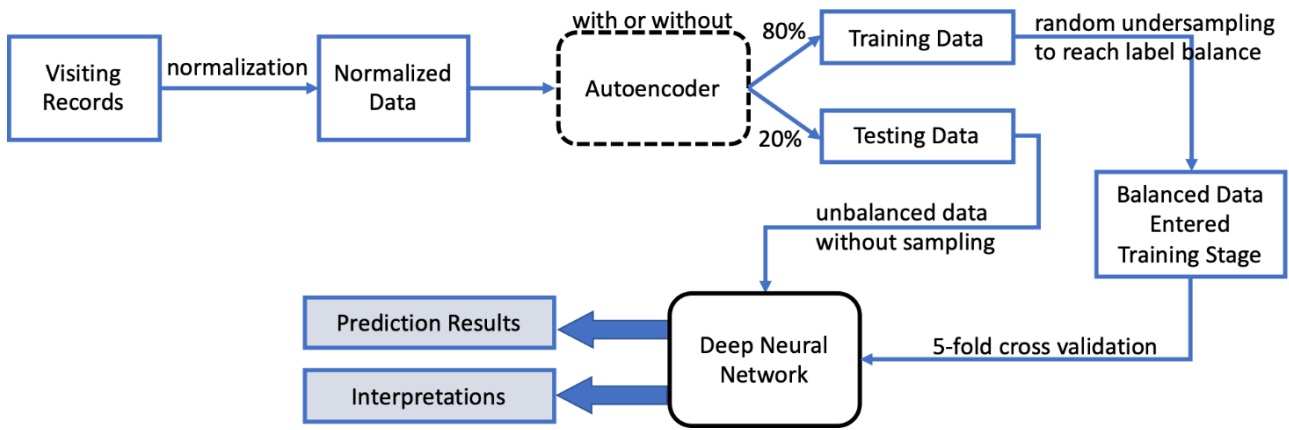

Figure 3. Data processing flow

The proposed DNN model contained 18 input nodes (based on input features) and 3 hidden layers with 100 neurons each. The rectified linear unit (ReLU) active functions were used for each layer. Four output nodes symbolized the 4 hospital levels. Optimization was carried out by mini-batch stochastic gradient descent that iterated through small subsets of the training data and modified the parameters in the opposite direction of the gradient of the loss function to minimize the reconstruction error. The data representation changing comparison is done through the same training process, except that one is preprocessed with AE, and the other is not, as shown in Figure 4. The AE model consisted of 5 hidden layers; the neurons in those layers were 500, 250, 100, 250, and 500. The ReLU active function was used for the encoder (first 2 layer) and decoder (last 2 layer) and a sigmoid active function for the latent space conversion (the third layer).

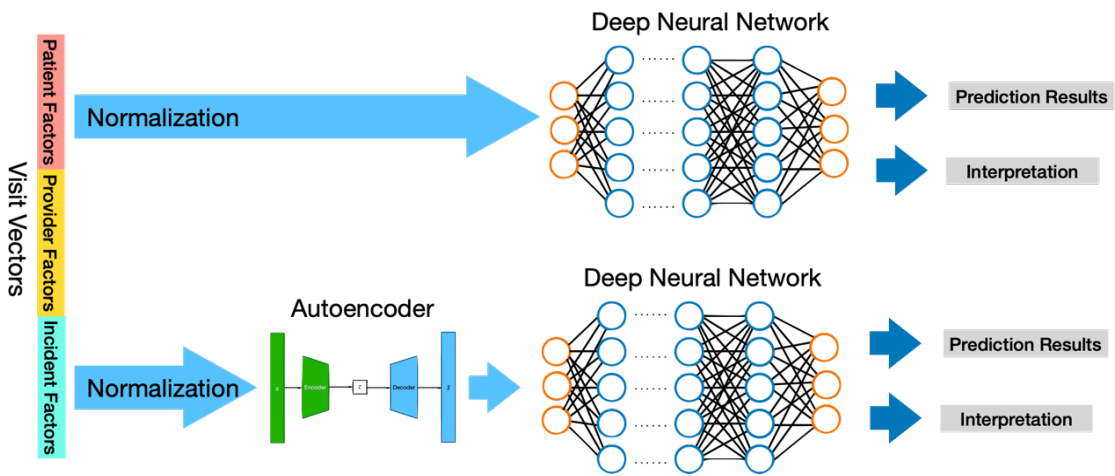

Figure 4. Comparing effectiveness of data representation changing

The performance indicators used here are the area under the receiver operating characteristics curve (AUC), accuracy, sensitivity/recall, specificity, precision, and F1 score, as shown in equations (11) to (15), where TP, TN, FP, and FN denote for true positive, true negative, false positive, and false negative. For a multi-class classification of hospital levels, the macro average was used to generalize the performance, which computed the metric independently for each class and then took the average to consider each class equally. The AUC used the one-vs-rest scheme to demonstrate the general performance. The result also compared the SHAP value of the trained model, with and without AE preprocessing, to show the effect of representation change on feature importance. This study was implemented with Python version 3.7.6, combined with PyTorch framework 1.1.0 and scikit-learn 0.22.2.

$$\text{Accuracy} = \frac{TP + TN}{TP + FP + FN + TN} \quad (11)$$

$$\text{Sensitivity} = \text{Recall} = \frac{TP}{TP + FN} \quad (12)$$

$$\text{Specificity} = \frac{TN}{TN + FP} \quad (13)$$

$$\text{Precision} = \frac{TP}{TP + FP} \quad (14)$$

$$F1 = \frac{2 * (\text{Precision} * \text{Recall})}{\text{Precision} + \text{Recall}} \quad (15)$$

## 3. Results

A total of 566,767 patients and 8,805,562 visiting records were analyzed, where 72.42% patients chose to go to the clinics, 8.35% to the district hospital, 10.73% to the regional hospital, and 8.51% to the medical center. Tables 1 to 3 demonstrate the information of patients, providers, and incidents. The performance results are listed in Table 4. Processing without and with AE reached an AUC of 0.87 and 0.90, respectively. Therefore, changing data representations led to an increase of 0.03 in AUC.

Table 1. Demographic information of patients

| Demographic information of patients (n =566,767) | |
|---|---:|
| **Age, mean (SD)** | 45.80 (16.33) |
| **Male, number of patients (%)** | 271,556 (47.91) |
| **Noted low income, number of patients (%)** | 12946 (2.28) |
| **Total number of diseases, mean (SD)** | 11.09 (7.92) |
| **Total number of chronic diseases, mean (SD)** | 11.00 (7.85) |
| **Total number of visits per patient, mean (SD)** | 16.70 (15.39) |
| **UPC, mean (SD)** | 0.52 (0.22) |
| **LUPC, mean (SD)** | 0.19 (0.25) |
| **COCI, mean (SD)** | 0.09 (0.17) |
| **SECOC, mean (SD)** | 0.43 (0.27) |

SD: Standard Deviation

Table 2. Information of Hospitals

| Hospital information (n = 42,697) | | |
|---|---|---|
| | | **Number of hospitals. (%)** |
| **Hospital levels** | **Medical center,** | 26 (0.06) |
| | **Regional hospital** | 131 (0.31) |
| | **District hospital** | 1,102 (2.58) |
| | **Clinic** | 41,438 (97.05) |
| | | **Mean (SD)** |
| **MFPC** | | 45.991 (168.76) |
| **LFPC** | | 44.005 (120.25) |
| **Physician density** | | 24.224 (21.90) |

SD: Standard Deviation

Table 3. Information of Incidents

| Information of Incidents (n = 8,805,562) | | |
|---|---|---|
| **Hospital levels, number of records (%)** | **Medical center,** | 749,293 (8.51) |
| | **Regional hospital** | 944,712 (10.73) |
| | **District hospital** | 734,976 (8.35) |
| | **Clinic** | 6,376,581 (72.42) |
| **Is surgery, number of records (%)** | | 245,908 (2.79) |
| **Is ER, number of records (%)** | | 159,265 (1.81) |
| **Is severe, number of records (%)** | | 307,563 (3.49) |
| **DIR, mean (SD)** | | 0.130 (0.14) |
| **Work day, number of records (%)** | | 7,373,122 (83.73) |

SD: Standard Deviation

Table 4. Performance Indicator of Choice of Hospital Level

|  | **Without AE preprocessing** | **With AE preprocessing** | **Increase** |
|---|---:|---:|---:|
| **AUC** | 0.87 | 0.90 | 0.03 |
| **Accuracy** | 0.81 | 0.90 | 0.09 |
| **F1 Score** | 0.82 | 0.88 | 0.06 |
| **Precision** | 0.82 | 0.86 | 0.04 |
| **Sensitivity** | 0.91 | 0.94 | 0.03 |
| **Specificity** | 0.94 | 0.97 | 0.03 |

The mean SHAP values are listed in Figure 5, with the MFPC, physician density, and LFPC listed as the top 3 factors. The contribution could be positive or negative. The contribution of other features, even combined together, was very less. The AE process changed the contribution ranking, shifting the physician density from the second to the third position. Figure 6 shows the interpretation analysis of individual cases. The base value represents the decision of the general public (56.13). The red color symbolizes the features that contributed positively and pushed the decision above the base value, and the blue color symbolizes the features that contributed negatively and pushed the decision below the base value. Figures 6(a) and 6(b) show cases processed by the model without AE, and 6(c) and 6(d) show cases processed by the model with AE. It appears that the processing of AE enlarges the less contributing features and makes them more visible on graphics.

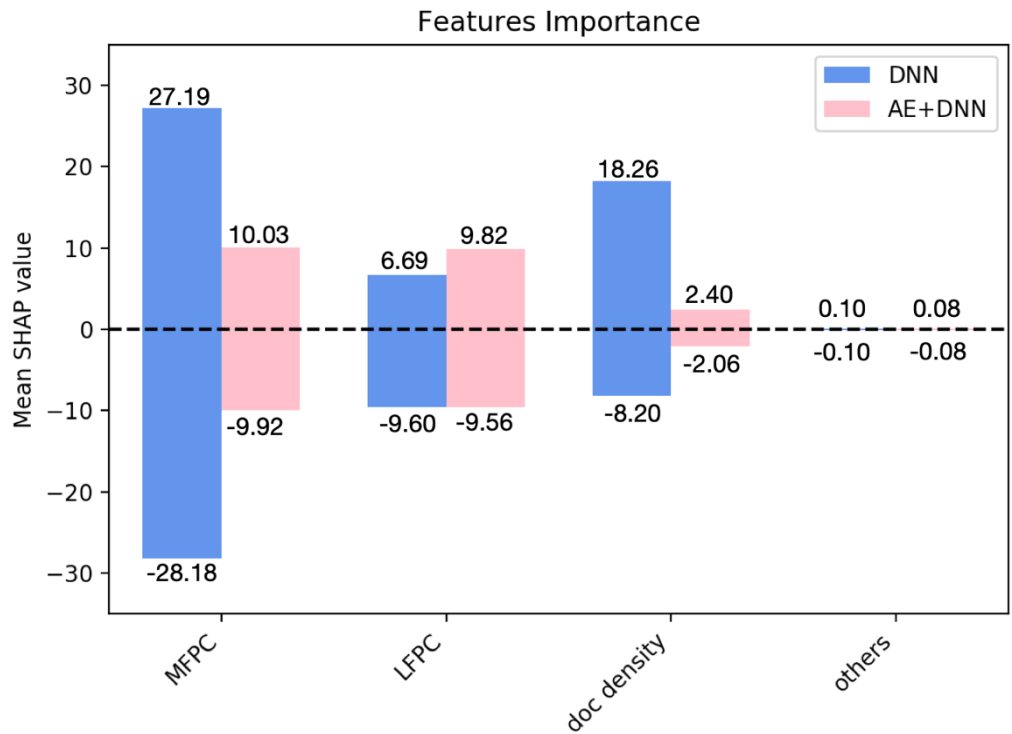

Figure 5. Model Interpretation with SHAP values

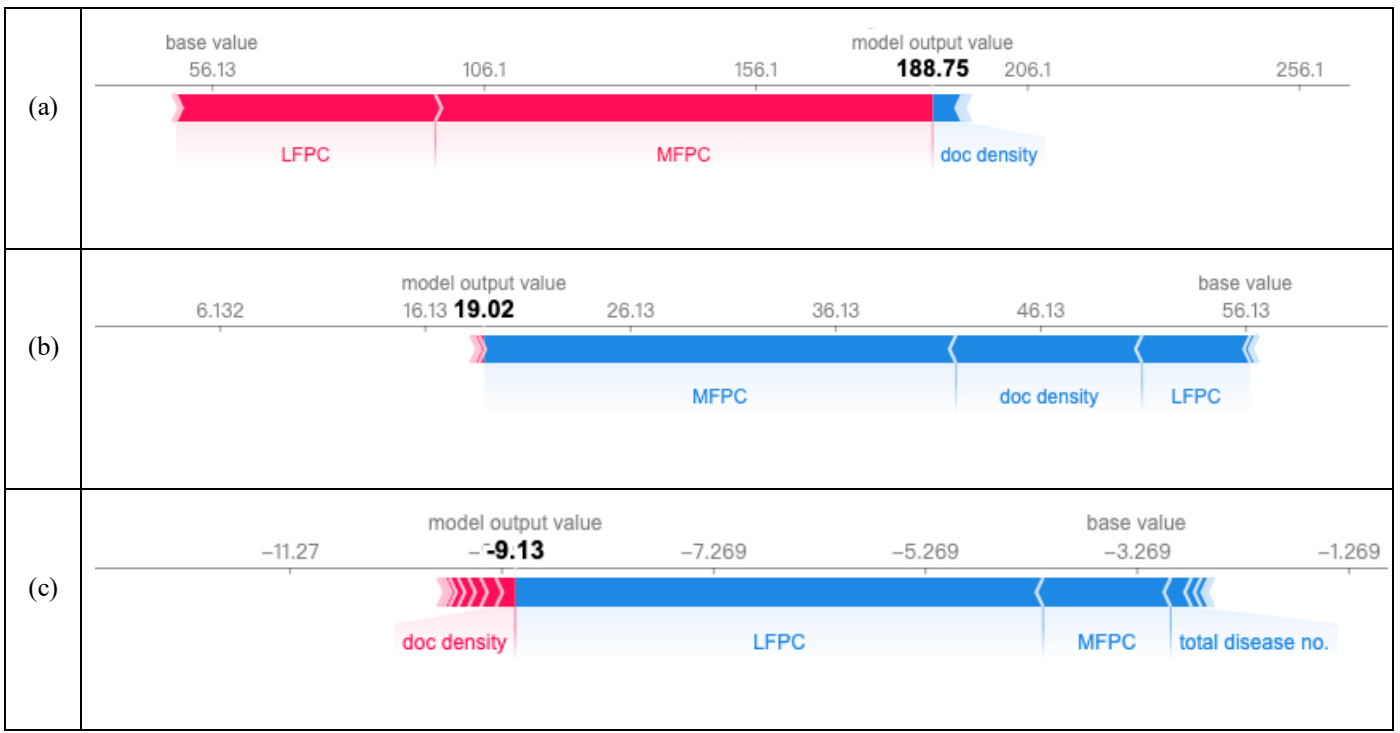

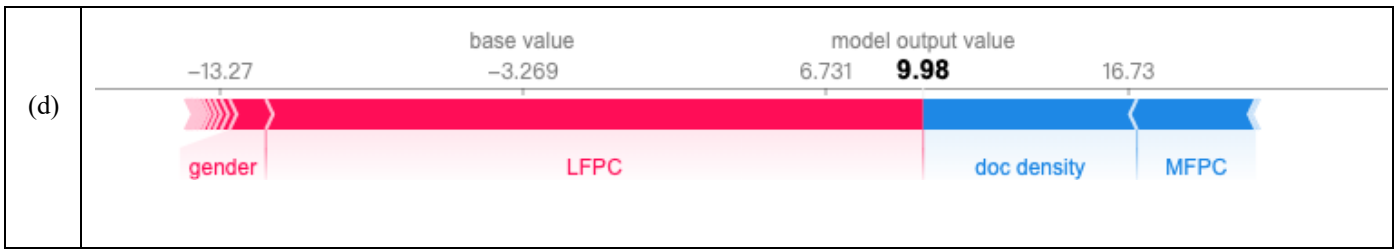

Figure 6. Interpretation analysis of individual cases. (a) and (b) are cases processed by the model without AE and (c) and (d) are cases processed by the model with AE.

## 4. Discussion

With the outbreak of COVID-19, it has been highlighted that there are insufficient applications in assisting public health. Exploring the patients' choices of hospital levels is the cornerstone of evaluating policies for referrals, utilization of healthcare resources, and rerouting patient volume for the hierarchization of services, and during disease outbreaks or infection control. The choice of patients was highly imbalanced. Some of the features did not reflect the character of the general public, such as low continuity, visits often not involving surgery, emergency service, or the condition not being considered severe, However, the model could still predict with high sensitivity and specificity, without initially finding the best-fit features. This study tried to address the black-box problem of machine learning [15, 47] using the SHAP value. According to our result, three features could interpret the majority of patients' choices of hospital levels: the MFPC, LFPC, and physician density. However, the features affected individuals differently.

Although entering the training process without previous data preprocessing is ideal, the performance of models can be improved by changing the data representation. It is straightforward that image and audio signals include disturbance and noise information, and using methods to eliminate noise leads to better predictions. Structured data were encoded with existing encoding schemes that were meaningful to people. The sparsity, discrete, and scarcity of data, which is invisible to the human eye, were difficult to notice. This study demonstrated the effectiveness of changing data representations. By merely adding preprocessing in advance, all the performance indicators increased, and the SHAP value became less extreme, allowing less contributing features to be observed.

The discipline of social economics mostly focused on clarifying the causality and the interrelationship of factors that affect patients in choosing hospitals. However, the machine learning approach attempted to seek the underlying pattern of data and predict accurately without relying on existing knowledge. The choice indicated a certain underlying trend, but not necessarily complete reasons and causalities. This study provided an alternative approach to observe the patients' choice. The prediction was based on the trajectory of the de-identified patient-visit data, commonly collected by the insurance company. Hence, the model is highly achievable elsewhere as it neither involves complex information that is difficult to collect nor violates patient privacy. However, this study has several limitations. Distance to travel remains an important factor in choosing hospitals [7]. Although there are ways of using de-identified insurance data to project the region where patients live [50], such information is still based on hypothesis and cannot be validated accurately. Further analysis could be done based on decisions of patients of different regions and explore different decision choices due to misallocation of medical resources. Future applications using deep learning technology are promising in health policy making. A golden standard for interpreting machine learning models has not been established. The method of using the interpretation model and the generated scores can be further explored.

**Acknowledgements**